# A short-key one-time pad cipher

Uwe Starossek

**Abstract**

A process for the secure transmission of data is presented that has to a certain degree the advantages of the one-time pad (OTP) cipher, that is, simplicity, speed, and information-theoretically security, but overcomes its fundamental weakness, the necessity of securely exchanging a key that is as long as the message. For each transmission, a dedicated one-time pad is generated for encrypting and decrypting the plaintext message. This one-time pad is built from a randomly chosen set of basic keys taken from a public library. Because the basic keys can be chosen and used multiple times, the method is called multiple-time pad (MTP) cipher. The information on the choice of basic keys is encoded in a short keyword that is transmitted by secure means. The process is made secure against known-plaintext attack by additional design elements. The process is particularly useful for high-speed transmission of mass data and video or audio streaming.

## 1 Introduction

A cipher is an algorithm for encryption and decryption with the purpose of securely transmitting data. The only information-theoretically secure and potentially unbreakable cipher is the one-time pad (OTP) cipher (U.S. Patent 1 310 719). It is simple to implement. Assuming the plaintext is a series of bits, it works as follows. A fresh OTP consisting of random series of bits that is at least as long as the plaintext is generated as a key. The plaintext is XORed with the key to create the ciphertext, meaning that each bit of the ciphertext is produced by an exclusive-or (XOR) logical operation applied to the corresponding two bits of plaintext and key. To decrypt the ciphertext, it is XORed with the same key. Because the same key is used for encryption and decryption, OTP is called a symmetric-key cipher.

Plaintext encrypted using OTP cannot be retrieved without the encrypting key. However, there are some conditions that must be met to not compromise the cipher. The practically most problematic one of these conditions is that the key must be transmitted to the receiver, but must not fall in the hands of a potential code breaker (in the following called attacker). This points to a fundamental weakness of the OTP cipher in that one must be able to securely transmit large amounts of data, that is, the key, to the receiver of the message.

When using other symmetric-key ciphers, such as the Advanced Encryption Standard (AES, NIST 2001), the problem of securely transmitting the symmetric-key cipher session key is often solved by using an asymmetric-key cipher, such as the Rivest Shamir Adleman (RSA) cipher, for key transfer.

In an asymmetric-key cipher, the public key of the receiver, which can be known to everyone, is used for encrypting messages and the private key of the receiver, which is known only to him, for decrypting messages. Public and private keys are mathematically related. While the public key can easily be computed from the private key, computation of the private key from the public key is possible in principle but requires an impractical amount of computational resources and time.

In RSA, the size of the public key is currently 1,024 to 4,096 bits. The size of the plaintext message must be smaller than the size of the key. Furthermore, randomized padding must be added to the message before encrypting it for ensuring security. This leaves the maximum RSA message size to be 117 bytes when the RSA key size is 1,024 bits (128 bytes). This is sufficient for transmitting the key of a symmetric-key cipher, such as AES with key sizes of 128 or 256 bits (16 or 32 bytes). However, it is insufficient for transmitting the key of the OTP cipher given that this key is as large as the message actually to be transmitted. Splitting the large OTP cipher key, or better directly the message, into 117-byte blocks and separately transmitting them with the RSA cipher is usually impractical because, compared to symmetric-key ciphers, this is too slow.

Hence it seems desirable to develop a process for the transmission of data that exhibits the advantages of the OTP cipher, namely simplicity, speed, and information-theoretically security, but overcomes its fundamental weakness, the necessity of securely exchanging a key that is as long as the message.

The Limited Storage Model proposed by Maurer [1], the Limited Access Model proposed by Rabin [2], and the Re-Randomizing Database concept proposed by Valiant [3] attempt to achieve this goal. All are based on the idea of publicly available random data. The Limited Storage Model requires an expensive random data distribution system, which tends to make it impractical. Its security relies on the supposedly high cost of data storage, which, considering the rapid decline of this cost, does not seem to be a permanently protecting factor. The Limited Access Model requires a voluntary network of a large number of computers each maintaining and updating random pages and acting as page server nodes, which impairs its practicality. Its security relies on the random data downloaded from the page server nodes not all being intercepted by an attacker, which makes it vulnerable. The Re-Randomizing Database concept requires the secure exchange of one key for each transmitted bit of information, which makes it impractical. Its security relies on the users' communication with the database not being intercepted by an attacker, which makes it vulnerable.

A hybrid cryptosystem having the desired properties is proposed here. A symmetric-key OTP-based cipher, in the following called multiple-time pad (MTP) cipher, is used for encrypting and decrypting the plaintext and another cipher, preferably an asymmetric-key cipher, is used for transmitting the symmetric key or, more precisely, information encoded in a short key from which the symmetric key can be computed. First, the basic design of the MTP cipher is outlined and its security against various kinds of attacks is discussed. It is then shown how a weakness to a particular kind of attack is overcome by an augmented design containing additional design elements. This outline is followed by a detailed description of the MTP cipher, a discussion of its security against various kinds of attack, and the description of variations of the MTP cipher.

## 2   Outline of the MTP cipher

A certain number $k$ of random binary basic keys of size $s$ are individually generated and consecutively numbered from 1 to $k$. The set of basic keys is known to sender and receiver. It can also be distributed publicly and be intended for general use thus serving for the exchange of messages between arbitrary parties. For maximum security, linear independency between the basic keys should be largely ensured. They are linearly independent if none of them can be obtained by XORing any number of the other basic keys. For the values of $k$ and $s$ outlined below, it is probable that this condition is sufficiently fulfilled provided the basic keys are generated in a perfectly random process. In the following, the basic keys are called MTP basic keys.

An alternative method is possible for generating the MTP basic keys as described in the next section. For the sake of simplicity, this description of the MTP cipher basic design is based on the method explained in the previous paragraph.

The sender of the message randomly chooses a subset of the set of $k$ MTP basic keys. The subset to be chosen has a size $g$, that is, it contains $g$ basic keys, where $g$ is in the range of 1 to $k$. The choice is defined by the set of serial numbers assigned to the chosen MTP basic keys. The sender then computes his private MTP key by performing a concatenated XOR operation on the chosen basic keys. That is, the first selected basic key is XORed with the second selected basic key, the result is XORed with the third selected basic key and so on until the result of the penultimate XOR operation is XORed with the $g^{th}$ selected basic key. Because XOR operations are commutative, the order of the individual operations in concatenated XOR operations can be changed without changing the result.

The sender uses his private MTP key for encrypting his binary plaintext message. The size of the message must be equal to or smaller than $s$, the size of both the MTP basic keys and the private MTP key. Encryption is done by XORing the plaintext with the private MTP key. The resulting MTP ciphertext is transmitted to the receiver.

In addition, the sender encodes, encrypts, and transmits to the receiver the set of serial numbers of the chosen MTP basic keys using an asymmetric-key cipher and the corresponding public key of the receiver. Because the existence or nonexistence of $k$ MTP basic keys in the private MTP key can be encoded in a string of $k$ bits, the encoding leads to a binary keyword of length $k$. The keyword ciphertext resulting from encrypting the keyword is transmitted to the receiver. Provided $k$ is properly chosen, the size of the data to be encrypted in this step is manageable by an asymmetric-key cipher such that it can be transmitted to the receiver in one unsplit transmission. If $k = 256$, for instance, the RSA cipher with a key size of 1,024 bits can be used because the maximum RSA message size then is 117 bytes and the size of the keyword is 256 bits, which correspondents to 32 bytes.

The receiver decrypts the keyword ciphertext using the asymmetric-key cipher and his private key and obtains the keyword from which he decodes the set of serial numbers of the selected MTP basic keys. With this information and the set of all numbered MTP basic keys, which are known to him, he can identify the MTP basic keys chosen by the sender and compute the sender's private MTP key by XORing the identified MTP basic keys.

Finally, the receiver XORs the MTP cyphertext with the sender's private MTP key and thus obtains the original binary plaintext message.

One advantage of the hybrid MTP cipher is that it inherits the simplicity of the OTP cipher. It can easily be implemented in computer software and hardware. Because it only involves the XOR command, which is executed in a native machine operation, it is fast. Thus, it is particularly appropriate for a secure transmission of mass data such as audio and video signals as well as for secure telephone communication. A further advantage is that the MTP basic keys need to be generated and distributed only once. The cost of ensuring randomness is amortized quickly.

The MTP cipher does not inherit the perfect information-theoretically security of the OTP cipher because the private MTP key space is smaller than the OTP key space. The security of the MTP cipher is further discussed in the following.

Ciphertext-only attacks are considered first. Vetting a new cipher includes testing of large quantities of ciphertext for any statistical departure from random noise. Assuming perfectly random MTP basic keys, the private MTP key generated from a randomly selected subset of MTP basic keys will also be perfectly random. When an arbitrary plaintext is XORed with such random key, the resulting ciphertext also is random. The same private MTP key should not be used twice to not compromise security. This is ensured with a high degree of probability by randomly choosing the MTP basic keys provided $k$ is large enough.

Another ciphertext-only attack is called brute-force attack, in which all possible keys are tried and the result of each decryption attempt is checked as to whether it is a valid plaintext. As shown further below, the size of the private MTP key space can easily be made large enough so that a brute-force attack will not succeed.

Next, a known-plaintext attack is considered, where the attacker knows the ciphertext and a part of the plaintext. It is assumed that $r$ bits of plaintext are known. When these bits are XORed with the corresponding bits of ciphertext, the corresponding bits of the private MTP key result. Based on this information, it can be attempted to identify the subset of $g$ chosen MTP basic keys, and thus the complete private MTP key, for retrieving the remaining plaintext. When $r = k$, this is an easy task because then a well-defined system of equations results that can be solved with Gauss's algorithm. Known plaintext of such a size is possible. In image or audio data, it can appear as long series of zeros. In such a case, the basic design of the MTP cipher described above is not secure against known-plaintext attack.

The weakness of the basic design of the MTP cipher to a known-plaintext attack is overcome by adding further design elements as described in the following.

In addition to XORing the plaintext with the private MTP key, it is XORed with a first random key of size $s$, which is separately generated by the sender for this purpose, and XORed with a second random key of size $s$. The second random key is obtained from the other keys, by a computation rule further specified in the next section. The resulting MTP ciphertext and the encrypted keyword carrying the information on the choice of MTP basic keys are transmitted to the receiver. Furthermore, the first random key is encrypted and the resulting MTP ciphertext is transmitted to the receiver using the same MTP cipher as used for the plaintext message, but a generally different private MTP key. The latter key is generated from a different subset of MTP basic keys, whose serial numbers are encoded in a keyword that is securely transmitted to the receiver using the asymmetric-key cipher, that is, as a second keyword ciphertext.

The receiver decrypts the two keyword ciphertexts using the asymmetric-key cipher and his private key and obtains the two keywords. From one keyword, he decodes the set of serial numbers of the MTP basic keys used for encrypting the plaintext. From the other keyword, he decodes the set of serial numbers of the MTP basic keys used for encrypting the first random key. With this information and the set of all numbered MTP basic keys, which are known to him, he computes the sender's private MTP keys used for encrypting the plaintext and the first random key.

The receiver XORs the encrypted plaintext, that is, the plaintext ciphertext, and the encrypted first random key, that is, the random key ciphertext, with the respective private MTP keys. From the first operation, the plaintext XORed with the first and second random keys results. From the second operation, the receiver obtains the first random key. He obtains the second random key by using the respective computation rule, which is known to him. The result of the first operation is XORed with the first and second random keys, which finally gives the original plaintext message.

The additional XORing of the plaintext with a random key aims at randomizing the plaintext so that a known-plaintext attack is averted. XORing with a second random key is provided to prevent the following scenario that is possible when only one random key is used. An attacker might know both the plaintext ciphertext and the random key ciphertext. When both ciphertexts are XORed, the random key cancels and the plaintext XORed with the sender's private MTP keys is obtained. The result would be almost as vulnerable to a known-plaintext attack as when using the basic design of the MTP cipher because the two remaining private MTP keys, when XORed, result in a combined private MTP key that has the same key space size as one single private MTP key. This vulnerability is eliminated by XORing with a second random key. Because the second random key is generated from the other keys, there is no need to transmit it to the receiver, which would compromise the cypher in the manner described above.

It is suggested to use an asymmetric-key cipher to transmit the information from which the private MTP keys can be deduced. Asymmetric-key ciphers in general and the RSA cipher in particular seem particularly suitable and preferable in the given context. Nevertheless, any sufficiently secure cipher, including symmetric-key ciphers and even the OTP cipher, can alternatively be used for transmitting the keywords that carry the private MTP key information. Further variations of the process outlined above are possible and described below.

## 3 Detailed description of the MTP cipher

### 3.1 XOR logical operation

The exclusive-or (XOR) logical operation used here is denoted by the symbol +. In

$$A = B + C \tag{1}$$

for example, the XOR operation is consecutively applied to two corresponding bits, that is, bits at corresponding positions, of *B* and *C* resulting in the corresponding bit of *A*, where *A*, *B*, and *C* are binary strings of the same length, that is, of the same number of bits. Because a binary string XORed with itself is a zero string (where all bits of the string are zero), that is,

$$C + C = 0 \tag{2}$$

and because XOR operations are commutative and the order of the individual operations in concatenated XOR operations does not affect the result, the above operation is inverted by applying it again, that is

$$B = A + C \tag{3}$$

### 3.2 MTP basic keys

The MTP basic keys are binary strings of length $s$. That is, each consists of a series of $s$ bits. The MTP basic keys can be generated in alternative ways.

In a first method, $k$ MTP basic keys $B_1$, $B_2$, …, $B_k$ are individually and independently generated. The MTP basic keys should be random and linearly independent with respect to XOR operations. They are available at both the sender and the receiver, possibly public, and intended for multiple use. Each key is assigned a different serial number between 1 and $k$. Linearly independent with respect to XOR operations means that none of the $k$ MTP basic keys follows from a single or concatenated XOR operation on two or more of the other MTP basic keys. Perfect linear independency is preferable, but a high degree of linear independency can be sufficient.

In a second method, the MTP basic keys are taken as substrings of a master string $B$ of length $l$, with $l \geq s$. Master string $B$ should be random and is available at both the sender and the receiver, possibly public, and intended for multiple use. An MTP basic key $B_q$ is identified by a pointer $q$ in the range $1 \leq q \leq l$. $B_q$ is taken as the substring of $B$ that commences at position $q$. When the remaining length of $B$ is shorter than the length of $B_q$, that is, when the sum of $q$ and $s$ is larger than $l$, the remaining final substring of $B_q$ is taken as the respective initial substring of $B$. This is equivalent to appending a copy of $B$ to itself or to connecting the last position of $B$ to its first position thus creating a looped string. In this method, $l$ different MTP basic keys are available. They are numbered consecutively from 1 to $l$, where the number that identifies an individual MTP basic key $B_q$ is the value of $q$ that indicates the position in $B$ of the first bit of $B_q$.

In both methods, the respective pre-generated data, that is, the set of MTP basic keys $B_1$, $B_2$, …, $B_k$ or the master string $B$, are called library. The randomness of this data should be as high as possible and preferably be perfect. Perfectly random data can be produced by physical random generators. The cost of ensuring perfect randomness is amortized quickly because the library is intended for multiple use and needs to be generated and distributed only once or, at least, only once in a certain period of use.

### 3.3 Private MTP keys

Two private MTP keys, $K_P$ and $K_R$, are generated. $K_P$ serves for encrypting the plaintext. $K_R$ serves for encrypting the first random key. They are generated by the following concatenated XOR operations

$$K_P = B_i + B_j + \ldots + B_m + B_n \tag{4}$$

$$K_R = B_r + B_s + \ldots + B_v + B_w \tag{5}$$

where $B_i, \ldots, B_n$ and $B_r, \ldots, B_w$ are MTP basic keys randomly chosen by the sender. In particular, $g$ out all available MTP basic keys are chosen, where also $g$ is generally random and generally differs for $K_P$ and $K_R$. The two private MTP keys are binary strings of length $s$.

### 3.4 First random key

The first random key, $R_1$, is a binary string of length $s$. For each application, it is freshly generated by the sender as a true or pseudo random number. Pseudo randomness can be sufficient depending on the kind of plaintext and the period of the pseudo random number sequence.

### 3.5 Second random key

The second random key, $R_2$, is a binary string of length $s$. It is generated from the other keys by a defined computation rule.

### 3.6 Computation rule

The computation rule defines how $R_2$ is generated from $R_1$ and/or $K_P$ and/or $K_R$. It should be such that both $R_2$ and $(R_1 + R_2)$ are as random as $R_1$ and different from $R_1$. The computation rule does not need to be invertible. For some variations of the MTP cipher, described below, it must not be invertible. Various computation rules are possible. For example, it can be a transposition rule. A transposition rule, for example, can be as follows. The first bit of $R_2$ is set to the second bit of $R_1$, the second bit of $R_2$ is set to the third bit of $R_1$, and so on until the penultimate bit of $R_2$ of is set to the last bit of $R_1$, and then the last bit of $R_2$ is set to the first bit of $R_1$. That is, all bits of $R_1$ are shifted one position up and the first bit of $R_1$ is moved to the end. Variations of this procedure are obtained by defining different relationships between the position of a respective bit of $R_2$ and the position of the corresponding bit of $R_1$ to which the respective bit of $R_2$ is set.

As a second example, the computation rule can be a transposition rule defined as follows. If the first bit of $K_P$ is zero, the first bit of $R_2$ is set to the second bit of $R_1$; otherwise, it is set to the third bit of $R_1$; if the second bit of $K_P$ is zero, the second bit of $R_2$ is set to the third bit of $R_1$; otherwise, it is set to the fourth bit of $R_1$; and so on until the penultimate bit of $R_2$ is set to the last or first bit of $R_1$ depending on whether the penultimate bit of $K_P$ is zero or one, and finally the last bit of $R_2$ is set to the first or second bit of $R_1$ depending on whether the last bit of $K_P$ is zero or one. Variations of this procedure are obtained by replacing $K_P$ by $K_R$ or some combination of $K_P$ and $K_R$, by exchanging $R_1$, $K_P$, and $K_R$ within such procedures, or by defining different relationships between the position of a respective bit of $R_2$ and the positions of the corresponding bits of $R_1$, $K_P$, and $K_R$ from which the respective bit of $R_2$ results.

### 3.7 Plaintext

The plaintext, $P$, is a binary string of length $s$. If the actual plaintext message is shorter than $s$ bits, the remaining bits of $P$ are set to zero. Alternatively, $P$, $K_P$, $K_R$, $R_1$, and $R_2$, as well as $C_P$ and $C_R$ defined below, are truncated to fit the length of the actual plaintext message.

### 3.8 Plaintext ciphertext

After having generated $K_P$, $R_1$, and $R_2$, the sender generates the plaintext ciphertext, $C_P$, by the following concatenated XOR operation

$$C_P = P + K_P + R_1 + R_2 \qquad (6)$$

$C_P$ is a binary string of length $s$. Because $R_1$ and $R_2$ are unknown to an attacker and because of the requirement that the computation rule for generating $R_2$ is such that $(R_1 + R_2)$ is as random as $R_1$, an attacker cannot deduce any information on the private MTP key $K_P$ even if $C_P$ and $P$ are partly or fully known. Hence a known-plaintext attack on $C_P$ will fail.

### 3.9 Random key ciphertext

After having generated $R_1$ and $K_R$, the sender generates the random key ciphertext, $C_R$, by the XOR operation

$$C_R = R_1 + K_R \qquad (7)$$

$C_R$ is a binary string of length $s$. Because $R_1$ is random, an attacker cannot gain any information from $C_R$ on $K_R$ or $R_1$.

### 3.10 Keywords

The sender generates two keywords, $W_P$ and $W_R$, in which the information on the MTP basic key choices are encoded. $W_P$ serves for identifying the MTP basic keys chosen for generating the private MTP key $K_P$. $W_R$ serves for identifying the MTP basic keys chosen for generating the private MTP key $K_R$.

If the MTP basic keys are generated by the first method described above and $g$ is not limited to small values (see below), the information on the choice of MTP basic keys is preferably encoded as follows. Each keyword is a binary string of length $k$. The bits of $W_P$ are set to one at the positions $i, j, \ldots, m, n$ (see Eq. (4)) and set to zero at all other positions. The bits of $W_R$ are set to one at the positions $r, s, \ldots, v, w$ (see Eq. (5)) and set to zero at all other positions. The serial numbers of the respectively chosen MTP basic keys are thus encoded in a compact form. When the MTP basic keys are generated by the first method and $g$ is limited to small values, the procedure outlined in the next paragraph, with $k$ instead of $l$, can also be advantageous.

If the MTP basic keys are generated by the second method described above, the information on the choice of MTP basic keys is preferably encoded as follows. Each of the respective pointer values $q_1, q_2, \ldots, q_g$ is encoded as a binary string. The individual strings are appended to each other thus forming the keyword $W_P$ or $W_R$, depending on whether the chosen MTP basic keys were used to generate $K_P$ or $K_R$. The binary strings representing the pointer values have a length $a$ that depends on $l$, the length of master string $B$. More specifically, $a$ is the logarithm of $l$ to base 2. Both keywords are binary strings of length $g \cdot a$, where $g$ is the number of the respectively chosen MTP basic keys.

### 3.11 Keyword ciphertexts

The sender encrypts the keywords $W_P$ and $W_R$ using an asymmetric-key cipher and the public key of the receiver. $W_P$ when encrypted gives the keyword ciphertext $A_P$. $W_R$ when encrypted gives the keyword ciphertext $A_R$. The keywords $W_P$ and $W_R$ could also be combined into a single keyword $W$, for instance, by appending $W_R$ to $W_P$, that is encrypted resulting in a combined keyword ciphertext $A$, instead of $A_P$ and $A_R$.

Alternatively, any sufficiently secure cypher, including symmetric-key cyphers, can be used for encrypting and securely transmitting the keywords.

### 3.12 Transmission

The sender transmits the plaintext ciphertext $C_P$, the random key ciphertext $C_R$, and the two keyword ciphertexts $A_P$ and $A_R$, or $A$, to the receiver.

The two ciphertexts $C_P$ and $C_R$ can be combined into one string, $C$, and transmitted as such (multiplex transmission). The rule of combination could be as follows: The bits of $C_P$ are consecutively represented by the odd-numbered bits, and the bits of $C_R$ are consecutively represented by the even-numbered bits, of the combined string $C$. Corresponding bits of $C_P$ and $C_R$ are then available at the receiver at virtually the same time. This is an advantage, in particular, when the data is streamed and to be continuously decrypted by the receiver. Also $A_P$ and $A_R$, or $A$, and $C$ can be combined into one string and transmitted as such so that the data transfer is entirely accomplished in a single overall combined string. For speed of decryption, the first bits of the combined string are those of $A_P$ and $A_R$, or $A$, followed by those of $C$.

### 3.13 Decryption

If the data have been combined before transmission, the receiver first decombines them. That is, he separates the received data to obtain the two ciphertexts $C_P$ and $C_R$ and the keyword ciphertexts $A_P$ and $A_R$ or the combined keyword ciphertext $A$.

Using the asymmetric-key cipher and his private key (or, alternatively, any sufficiently secure cypher), the receiver decrypts the two keywords $W_P$ and $W_R$ from the keyword ciphertexts $A_P$ and $A_R$, or he decrypts the combined keyword $W$ from the combined keyword ciphertext $A$ and then separates $W_P$ and $W_R$ from $W$.

From $W_P$, he decodes the set of serial numbers or pointer values that identify the MTP basic keys used for generating $K_P$. From $W_R$, he decodes the set of serial numbers or pointer values that identify the MTP basic keys used for generating $K_R$. Knowing the entire set of $k$ MTP basic keys or the master string $B$ of length $l$, he computes $K_P$ and $K_R$ using Eqs. (4) and (5). The receiver then computes the first random key by the XOR operation

$$R_1 = C_R + K_R \qquad (8)$$

which, according to Eqs. (1) and (3), is the inversion of operation (7). Knowing $R_1$ and, if required, $K_P$ and $K_R$, he obtains the second random key, $R_2$, by using the known computation rule. In the final step, the receiver finds the plaintext from the concatenated XOR operation

$$P = C_P + K_P + R_1 + R_2 \tag{9}$$

which is the inversion of operation (6).

## 4 Security

### 4.1 Brute-force attack

Security against brute-force attack, that is, an exhaustive key search, is considered separately for the two methods of generating the MTP basic keys described above.

If the MTP basic keys are generated by the first method described above and $g$ is random with $g \leq k$, the size of the private MTP key space, that is, the number of different such keys, corresponds to $2^k$. Assuming that the number of MTP basic keys is $k = 256$, the number of private MTP keys is $2^{256}$. This is much larger than $2^{128}$, the key space size that, based on a physical argument and the Landauer limit, is considered computationally secure if the keys are used in a symmetric-key cipher.

If the MTP basic keys are generated by the second method described above, the size of the private MTP key space is different. For instance, when the length of the master string $B$ is $l = 2^{32}$, corresponding to a size of $B$ of 512 MB, and $g$ is set to $g = 4$ (instead of being random with $g \leq l$), the number of private MTP keys is almost $(2^{32})^4 = 2^{128}$ which is computationally secure. Apparently, $g$ can be comparatively small in this variation and the computational effort in evaluating Eqs. (4) and (5) is correspondingly reduced.

The above considerations refer to the case that only one private MTP key needs to be found. When using the cipher described by Eqs. (6) to (9), however, two keys must actually be found. Using the same parameters, the combined key space size then is $(2^{256})^2 = 2^{512}$ in the first method and $(2^{128})^2 = 2^{256}$ in the second method. Hence $k$ could be reduced to $k = 128$ or smaller in the first method (at the same time keeping $g$ random with $g \leq k$) and $g$ could be reduced to $g = 2$ in the second method.

In summary, the MTP cipher is secure against brute-force attack provided parameter $k$ or $l$ on the one side and parameter $g$ on the other are properly chosen.

Care must be taken that the choice of the subset of MTP basic keys defining a private MTP key is perfectly random so that each subset within the limits defined for $g$ (see below) has the same chance of being chosen.

### 4.2 Known-plaintext attack

It was shown above that a known-plaintext attack will fail when only $C_P$ and $P$ are partly or fully known. If an attacker knows both the plaintext ciphertext $C_P$ and the random key ciphertext $C_R$, he can XOR both strings leading to

$$C_P + C_R = (P + K_P + R_1 + R_2) + (R_1 + K_R) = P + K_P + K_R + R_2 \tag{10}$$

Although $R_1$ has canceled, $R_2$ is still present in the result. Thus, even if $P$ is partly or fully known, for instance, if $P$ is assumed to be a zero string, no information on the private MTP keys $K_P$ and $K_R$ can be gained from this operation because these keys are still randomized by the presence of $R_2$ that is unknown to the attacker. Hence the MTP cipher is secure against known-plaintext attack.

## 5   Parameter choice

The parameters $s$, $k$ or $l$, and $g$ are chosen such that the speed of data transfer and the computational hardware requirements are reasonably balanced. Furthermore, the parameters $k$ or $l$, and $g$ are chosen such that a brute-force attack will fail.

If the MTP basic keys are generated by the first method described above, the $k$ MTP basic keys must be stored by sender and receiver. The corresponding data volume is $k \cdot s$. For the computational hardware available today, a reasonable choice of $s$ could be in the range of $2^{23}$ to $2^{30}$ (corresponding to 1 MB to 128 MB), depending on the sort of data to be transferred. If $k = 256$, for example, the respective data volume to be permanently stored would be 256 MB to 32 GB.

If the MTP basic keys are generated by the second method described above, only the master string $B$ must be stored by sender and receiver. If the size of $B$ is chosen to be $l = 2^{32}$, for example, the respective data volume to be permanently stored would be 512 MB. This volume is independent of $s$, which could be as large as $l$.

The computational effort in evaluating Eqs. (4) and (5) depends on the size $s$ of the MTP basic keys and on the number $g$ of MTP basic keys chosen to generate the private MTP keys.

If the MTP basic keys are generated by the first method, $g$ is generally in the range of 1 to $k$. However, it can be advantageous to limit $g$ to minimum and maximum values, $g_{min}$ and $g_{max}$, so that $g_{min} \leq g \leq g_{max}$. Choosing a maximum value $g_{max}$ that is smaller than $k$ has the advantage of reducing the effort required for computing the private MTP keys. At the same time, this would reduce the size of the private MTP key space and hence the security against brute-force attack. When the number of MTP basic keys is $k = 256$ and $g$ is set to $g = g_{min} = g_{max} = 16$, for instance, the number of different private MTP keys is $b(256|16) \approx 1.033 \cdot 2^{83}$, where $b(k|g)$ is the binomial coefficient indexed by $k$ and $g$. The combined key space of two private MTP keys then has a size of $(b(256|16))^2 \approx 1.068 \cdot 2^{166}$. This is larger than $2^{128}$, the key space size considered computationally secure (see above). Hence allowing $g$ to be in the range $0 \leq g \leq 16$ would also be computationally secure. In the latter case, the average computational effort in evaluating Eqs. (4) and (5) would be reduced by a factor of 16 compared to allowing the maximum possible range of $0 \leq g \leq 256$.

If the MTP basic keys are generated by the second method, $g$ can be set to a comparatively small value as shown in the above discussion of brute-force attack. The computational effort in evaluating Eqs. (4) and (5) is correspondingly small.

When the keywords are transmitted using an asymmetric-key cipher, they should not be too long.

When using the first encoding method described above, the keyword $W_P$ and $W_R$ are each binary strings of length $k$. If $k = 256$, for example, the size of one keyword is 256 bits, or 32 bytes, and the combined size of two keywords is 64 bytes. Hence the RSA cipher with a key size of 1,024 bits can be used to transmit both keywords in one unsplit transmission given that the maximum RSA message size then is 117 bytes.

When using the second encoding method described above, the keyword $W_P$ and $W_R$ are each binary strings of length $g \cdot a$. If the MTP basic keys are taken as substrings of a master string $B$ of length $l$ that is chosen to be $l = 2^{32}$ and $g$ is set to $g = 2$ (reasonable choices with respect to security against brute-force attack, see above), for example, then $a = \log_2(l) = 32$ and $g \cdot a = 2 \cdot 32 = 64$. Hence the size of one keyword is 64 bits, or 8 bytes, and the combined size of two keywords is 16 bytes. Again, this is smaller than 117 bytes and both keywords can be transmitted to the receiver in one unsplit transmission.

# 6  Advantages

The MTP cipher partly inherits the simplicity of the OTP cipher and can easily be implemented in computer software and hardware. Its advantage over the OTP cipher is that the key that is to be transmitted to the receiver is much shorter. Its advantage over other processes that involve the OTP cipher [1, 2, 3] is that the required random data is produced and handled more easily.

A further advantage is that, once the private MTP keys $K_P$ and $K_R$ are generated by the sender and the information from which they can be regenerated is transmitted to the receiver, the ciphertexts $C_P$ and $C_R$ can continuously be generated by the sender and streamed to the receiver, who can continuously decrypt the data and produce the plaintext. That is, the receiver can begin to play the plaintext data, such as an audio or video, before the entire plaintext message has been encrypted and transmitted by the sender.

For this purpose, the computation rule that generates $R_2$ from $R_1$ and/or $K_P$ and/or $K_R$ must be such that also $R_2$ can be generated continuously. The computation rules using transposition defined above fulfill this requirement for the MTP cipher detailed in Section 3, but not necessarily for the variations of this cipher described below.

# 7  Variations

The MTP cipher can be varied in various ways leading to alternative processes that require the same or less degree of computational effort and provide the same or less degree of security. According to Eq. (6), the plaintext $P$ is XORed with three different keys: the private MTP key $K_P$, the first random key $R_1$, and the second random key $R_2$. Variations are obtained by replacing Eq. (6) with equations in which $P$ is XORed with any one or any two of these three keys. Further variations are obtained by replacing Eq. (7) with equations in which the second private key $K_R$ is not XORed with the first random key $R_1$ but with the second random key $R_2$ or with both the first random key $R_1$ and the second random key $R_2$. This leads to 21 different combinations of Eqs. (6) and (7) and their respective substitutes. Equations (8) and (9) are to be adapted accordingly. However, not all of these combinations are feasible and secure. In the following, some feasible variations are discussed.

For example, the second random key $R_2$ could be included in the generation of the random key ciphertext $C_R$ instead of in the plaintext ciphertext $C_P$. This is equivalent to replacing the operations described by Eqs. (6) to (9) by the following operations:

$$C_P = P + K_P + R_1 \tag{6a}$$

$$C_R = R_1 + R_2 + K_R \tag{7a}$$

$$R_1 + R_2 = C_R + K_R \tag{8a}$$

$$P = C_P + K_P + R_1 \tag{9a}$$

Before performing operation (9a), $R_1$ would have to be determined from $(R_1 + R_2)$ and possibly $K_P$ and $K_R$. The computation rule for generating $R_2$ would have to be such that this is possible. When the first of the two computation rules described above has been used for generating $R_2$, the first bit of the string to be calculated would have to be set to 0 or 1 leading to a string that could be either $R_1$ or $\neg R_1$ (where $\neg$ denotes an operator that flips all bits of the string it precedes). Hence both possibilities would have to be checked by using the resulting string for $R_1$ in operation (9a).

Another variation of the MTP cipher is obtained by omitting the private MTP key $K_P$. In this case, the following set of operations is used instead of the one described by Eqs. (6) to (9):

$$C_P = P + R_1 + R_2 \tag{6b}$$

$$C_R = R_1 + K_R \tag{7b}$$

$$R_1 = C_R + K_R \tag{8b}$$

$$P = C_P + R_1 + R_2 \tag{9b}$$

This variation can be vulnerable to a known-plaintext attack. Assume an attacker knows both ciphertexts, $C_P$ and $C_R$, and a part of the plaintext $P$. He can then determine the corresponding bits of $(R_1 + R_2)$ from inverting operation (6b). Depending on the computation rule for generating $R_2$, bits of $R_1$ might be deducible from bits of $(R_1 + R_2)$ and the corresponding bits of $K_R$ then follow from inverting operation (7b). If enough bits of $K_R$ are known, the full private MTP key $K_R$ can be reconstructed (as explained above) and performing operation (8b), then applying the computation rule, and finally performing operation (9b) would disclose the plaintext. Thus, the computation rule would have to be such that bits of $R_1$ are not easily deducible from bits of $(R_1 + R_2)$. This can be achieved by using a computation rule that includes one or both of the private MTP keys, such as the second transposition rule described above. Alternatively, care must be taken that an attacker cannot obtain knowledge of, or guess, too many bits of plaintext. Because only one private MTP key is present, the parameters (i.e., $k$, $l$, $g$) must be chosen greater to achieve the same degree of security against brute-force attack. Hence omitting the private MTP key $K_P$ does not much reduce the overall computational effort.

A variation of the previous variation is obtained by omitting the first random key $R_1$, in addition to the private MTP key $K_P$, when generating the plaintext ciphertext $C_P$. In this case, the following set of operations is used instead of the one described by Eqs. (6) to (9):

$$C_P = P + R_2 \tag{6c}$$

$$C_R = R_1 + K_R \tag{7c}$$

$$R_1 = C_R + K_R \tag{8c}$$

$$P = C_P + R_2 \tag{9c}$$

With a similar reasoning as for the previous variation, it can be shown that this variation is vulnerable to a known-plaintext attack, in this case, when bits of $R_1$ are easily deducible from bits of $R_2$. And, as before, the parameters must be chosen greater to achieve the same degree of security against brute-force attack.

Another variation of the MTP cipher is obtained by omitting only the first random key $R_1$, but keeping the private MTP key $K_P$, when generating the plaintext ciphertext $C_P$. In this case, the following set of operations is used instead of the one described by Eqs. (6) to (9):

$$C_P = P + K_P + R_2 \tag{6d}$$

$$C_R = R_1 + K_R \tag{7d}$$

$$R_1 = C_R + K_R \tag{8d}$$

$$P = C_P + K_P + R_2 \tag{9d}$$

Instead of by Eq. (10), the result of XORing $C_P$ and $C_R$ would be described by

$$C_P + C_R = (P + K_P + R_2) + (R_1 + K_R) = P + K_P + K_R + R_1 + R_2 \tag{10d}$$

Both $R_1$ and $R_2$ remain present in the result. Because of the initial requirement that the computation rule for generating $R_2$ is such that $(R_1 + R_2)$ is as random as $R_1$, an attacker cannot gain any information on the private MTP keys $K_P$ and $K_R$ even if $P$ is partly or fully known. However, if an attacker appropriately transposes $C_P$ before XORing it with $C_R$, the random keys $R_1$ and $R_2$ may cancel in case of certain choices of the computation rule, in particular, the first computation rule described above. Hence the computation rule must be carefully chosen to not make this variation vulnerable to a known-plaintext attack.

Another variation of the MTP cipher is obtained by using only one and the same private MTP key, $K$, for producing both the plaintext ciphertext $C_P$ and the random key ciphertext $C_R$. The following set of operations is then used instead of the one described by Eqs. (4) to (9):

$$K = B_i + B_j + \ldots + B_m + B_n \tag{4e}$$

$$C_P = P + K + R_1 + R_2 \tag{6e}$$

$$C_R = R_1 + K \tag{7e}$$

$$R_1 = C_R + K \tag{8e}$$

$$P = C_P + K + R_1 + R_2 \tag{9e}$$

This variation can be vulnerable to a known-plaintext attack. Consider that the result of XORing $C_P$ and $C_R$ would be described by

$$C_P + C_R = (P + K + R_1 + R_2) + (R_1 + K) = P + R_2 \tag{10e}$$

If the plaintext $P$ is partially known, the corresponding bits of $R_2$ can be determined from an inverted Eq. (10e). Depending on the computation rule for generating $R_2$, bits of $R_1$ might be deducible from $R_2$ and bits of $K$ would then follow from an inverted Eq. (7e). If enough bits of $K$ are known, the full private MTP key $K$ can be reconstructed. $R_1$ would then follow from Eq. (8e), $R_2$ from applying the computation rule, and finally the plaintext $P$ from performing Eq. (9e). Thus, the computation rule would have to be such that bits of $R_1$ are not easily deducible from bits of $R_2$. Again, this can be achieved by using a computation rule that includes one or both of the private MTP keys, such as the second transposition rule described above. Alternatively, care must be taken that an attacker cannot obtain knowledge of, or guess, too many bits of plaintext. Also here, the parameters (i.e., $k$, $l$, $g$) must be chosen greater to achieve the same degree of security against brute-force attack as achieved when using two private MTP keys so that using only one private MTP key does not much reduce the overall computational effort.

The basic design of the MTP cipher outlined at the beginning of Section 2 can be considered another, simplified, variation of the augmented design of the MTP cipher detailed in Section 3. It is characterized in that the two random keys $R_1$ and $R_2$ and the second private MTP key $K_R$ are omitted in the process. Thus, the following set of operations is used instead of the one described by Eqs. (6) to (9):

$$C_P = P + K_P \tag{6f}$$

$$P = C_P + K_P \tag{9f}$$

As shown in Section 2, this variation is vulnerable to known-plaintext attack. Nevertheless, it might be sufficiently secure provided an attacker cannot obtain knowledge of, or guess, too many bits of plaintext. Again, because only one private MTP key is present, the parameters must be chosen greater to achieve the same degree of security against brute-force attack.

Further variations of the MTP cipher are produced by replacing the XOR operations with other logical operations such as exclusive-nor (XNOR) logical operations.

## 8  Conclusions

A cipher for the secure and rapid transmission of data has been presented that is particularly useful for the high-speed transmission of mass data and for video or audio streaming. Its security and speed stems from encrypting and decrypting a plaintext with a one-time pad (OTP) key that is generated from a random choice of random basic keys possibly kept in a public library. Because the basic keys can be chosen and used multiple times, the method is called multiple-time pad (MTP) cipher. The information on the respective choice of basic keys is encoded in a short keyword that is transmitted to the receiver by secure means, preferably via an asymmetric-key cipher, so that the receiver can regenerate the OTP used by the sender and decrypt the plaintext.

It has been shown that the cipher is secure against brute-force attack provided the parameters are properly chosen. Security against known-plaintext attack is provided by encrypting and decrypting the plaintext with additional keys, that is, a first random key freshly generated by the sender for each application, and a second random key generated by the sender from the other keys by using a defined computation rule. The first random key is securely transmitted using the same MTP method as for transmitting the plaintext. Security against known-plaintext attack stems mainly from the second random key not being transmitted but regenerated by the receiver from the other keys using the computation rule.